\documentclass[preprint,12pt]{elsarticle}

\usepackage{xcolor}
\usepackage{times}
\usepackage{epsfig}
\usepackage{graphicx}
\usepackage{amsmath}
\usepackage{amssymb}
\usepackage{mathtools}
\usepackage{multirow}
\usepackage{comment}

\let\given\givenbase

\usepackage{algorithm}
\usepackage{algorithmic}
\usepackage{subfigure}

\def\revised{\textcolor{black}}

\usepackage{soul}
\setstcolor{red}

\usepackage{lineno,hyperref}

\journal{Applied Soft Computing}

\begin{document}
\sloppy

\begin{frontmatter}

\title{Robust Weakly Supervised Learning for COVID-19 Recognition Using Multi-Center CT Images}

\author[label1,label2]{Qinghao Ye\corref{cor1}}
\cortext[cor1]{These authors contributed equally to this work.}
\author[label3,label4]{Yuan Gao\corref{cor1}}
\author[label33]{Weiping Ding\corref{cor1}}
\author[label4]{Zhangming Niu}
\author[label5]{Chengjia Wang}
\author[label1,label6]{Yinghui Jiang}
\author[label1,label6]{Minhao Wang}
\author[label7]{Evandro Fei Fang}
\author[label4]{Wade Menpes-Smith}
\author[label8]{Jun Xia\corref{cor2}}
\ead{xiajun@email.szu.edu.cn}
\author[label9,label10]{Guang Yang\corref{cor2}}
\cortext[cor2]{Corresponding authors.}
\ead{g.yang@imperial.ac.uk}

\address[label1]{Hangzhou Ocean's Smart Boya Co., Ltd}
\address[label2]{University of California, San Diego, La Jolla, California, USA}
\address[label3]{Institute of Biomedical Engineering, University of Oxford, UK}
\address[label4]{Aladdin Healthcare Technologies Ltd}
\address[label33]{Nantong University, Nantong 226019, China}
\address[label5]{BHF Center for Cardiovascular Science, University of Edinburgh, Edinburgh, UK}
\address[label6]{Mind Rank Ltd}
\address[label7]{Department of Clinical Molecular Biology, University of Oslo, Norway}
\address[label8]{Radiology Department, Shenzhen Second People’s Hospital, Shenzhen, China}
\address[label9]{Royal Brompton Hospital, London, UK}
\address[label10]{National Heart and Lung Institute, Imperial College London, London, UK}





\begin{abstract}
\sloppy
The world is currently experiencing an ongoing pandemic of an infectious disease named coronavirus disease 2019 (i.e., COVID-19), which is caused by the severe acute respiratory syndrome coronavirus 2 (SARS-CoV-2). Computed Tomography (CT) plays an important role in assessing the severity of the infection and can also be used to identify those symptomatic and asymptomatic COVID-19 carriers. With a surge of the cumulative number of COVID-19 patients, radiologists are increasingly stressed to examine the CT scans manually. Therefore, an automated 3D CT scan recognition tool is highly in demand since the manual analysis is time-consuming for radiologists and their fatigue can cause possible misjudgment. However, due to various technical specifications of CT scanners located in different hospitals, the appearance of CT images can be significantly different leading to the failure of many automated image recognition approaches. The multi-domain shift problem for the multi-center and multi-scanner studies is therefore nontrivial that is also crucial for a dependable recognition and critical for reproducible and objective diagnosis and prognosis. In this paper, we proposed a COVID-19 CT scan recognition model namely coronavirus information fusion and diagnosis network (CIFD-Net) that can efficiently handle the multi-domain shift problem via a new robust weakly supervised learning paradigm. Our model can resolve the problem of different appearance in CT scan images reliably and efficiently while attaining higher accuracy compared to other state-of-the-art methods.
\end{abstract}


\end{frontmatter}


\section{Introduction}

The pandemic of coronavirus disease (COVID-19) is spreading all over the world rapidly. The number of infections is growing exponentially in different regions, which has triggered great health concerns in the international communities. One of the effective diagnostic methods confirmed by the World Health Organization is via viral nucleic acid detection using the reverse transcription polymerase chain reaction (RT-PCR) test \cite{wang2020deep}. However, the RT-PCR test is not sensitive sufficiently in some cases, which may put hurdles for presumptive patients to be identified and treated early.

As a non-invasive imaging technique, computed tomography (CT) can detect those characteristics, e.g., bilateral patchy shadows or ground glass opacity (GGO), manifested commonly in the COVID-19 infected lung. Hence CT may serve as an important tool for COVID-19 patients to be pre-screened and diagnosed early. The quantified imaging biomarkers extracted from CT images can also provide crucial prognostic values. Recently, deep learning based methods have been developed efficiently for the chest X-ray/CT data analysis and classification \cite{li2020artificial, wang2020covid, ouyang2020dual}, and these approaches can achieve state-of-the-art performance on X-ray/CT image diagnosis and prognosis.

\begin{figure}
\centering
\includegraphics[width=0.58\linewidth]{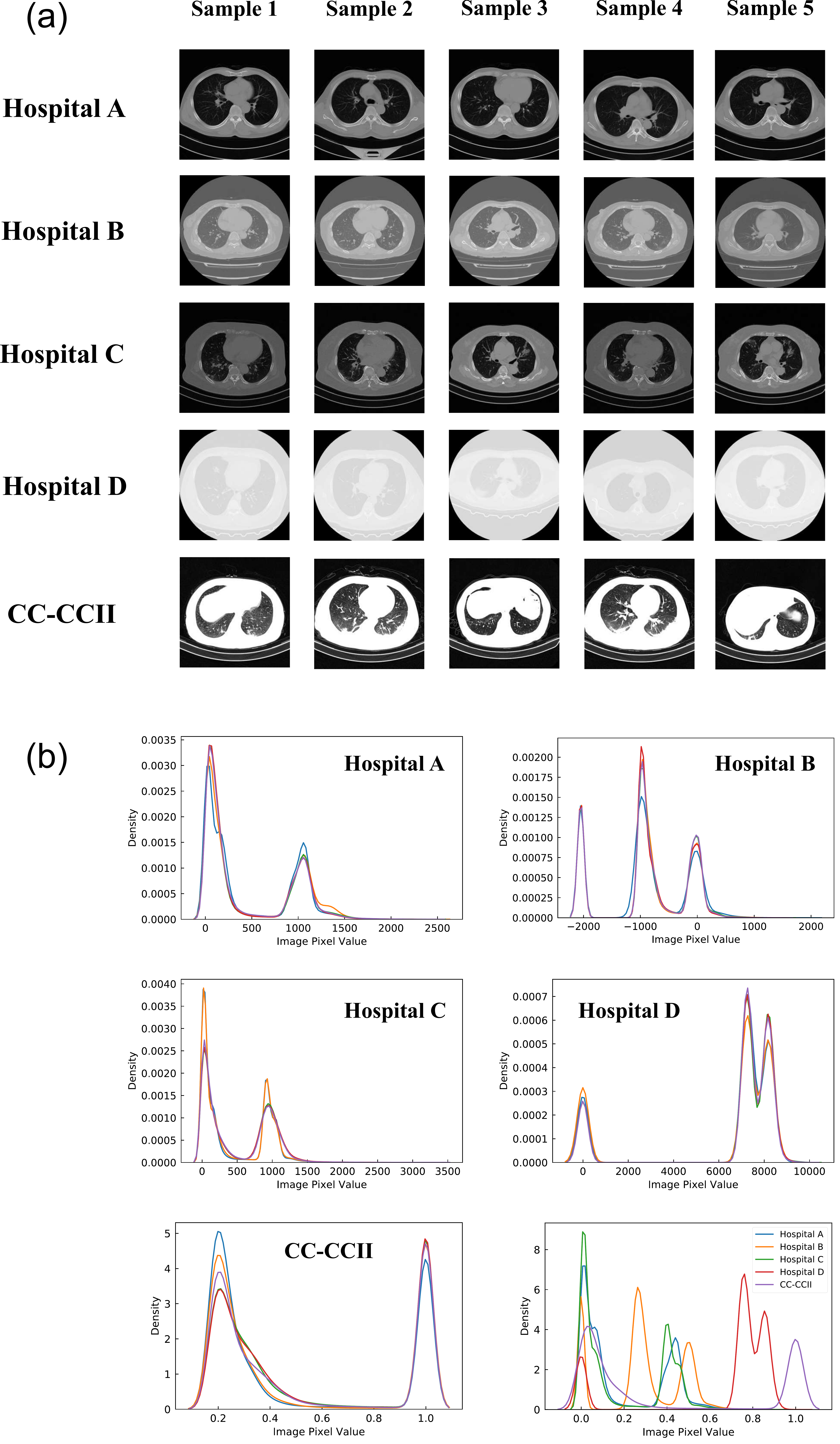}
\caption{(a) Samples of CT images are taken from five different hospitals and (b) The histograms of these CT images. Compared with images from Hospital A and Hospital D, it is clear that the brightness levels are distinctive. Moreover, the contrast of the data collected from the China Consortium of Chest CT Image Investigation (CC-CCII) dataset is considerably different from CT images acquired from other hospitals. The right bottom figure demonstrates the distribution of the images from different hospitals after normalization, however, these distributions still behave distinctively. It is of note that there are no visually distinctive features across CT scan images but it is easy for human radiologists to correctly classify despite CT scanner changes. On the contrary, deep learning based automated methods may fail to generalize across CT images acquired from different hospitals.}
\label{fig:sample}
\end{figure}

Nevertheless, most CT scan datasets for COVID-19 only contain CT volumes with a set of CT slices with only patient-level annotations provided (i.e., patient-level class labels available) indicating the patient is infected or not. There is a lack of per-slice labels since annotating each slice is labor-intensive and time-consuming for radiologists. It has been reported that it could take  an experienced radiologist about 21.5 minutes \cite{lin2005emergency} to analyze and label one whole CT volume. Consequently, convolutional neural network (CNN) based deep learning models trained on CT slices with only the patient-level label can perform poorly because some annotations of these CT slices are incorrect (e.g., non-lesion slices of the lung are actually be falsely labeled) leading training data to be noisy.

Yet another challenge when employing deep learning methods to medical image recognition is called \textit{data distribution shift} (a.k.a., \textbf{multi-domain shift}). Data distribution shift refers to the phenomenon that a common object or organ collected under various scenarios (e.g., different machine vendors and  sequence parameters) can result in vastly different data distributions. Therefore, models trained under the empirical risk minimization (ERM) \cite{vapnik1992principles} might cause the failure of model generalization. It is because the ERM assumes that training and testing data are sampled from the same or similar distribution and domains. However, in the data distribution shift scenario, this assumption would be violated. 

When a neural network is trained with images from one domain and tested on another domain (i.e., distinct imaging scenarios), the recognition performance often degrades dramatically. Figure \ref{fig:sample} represents images of different CT data collected from different hospitals. In the figure, it can be observed that CT data obtained from different hospitals are visually different although they all present image slices of the lung. It is on the grounds that every hospital uses different protocols and parameters for CT scanners when collecting the images for patients. Therefore, the multi-domain shift problem of the multi-center and multi-scanner studies is nontrivial. It is crucial to solving the multi-domain shift problem to achieve a dependable recognition, which is critical for reproducible diagnosis and prognosis.

\begin{figure*}
\centering
\includegraphics[width=0.9\linewidth]{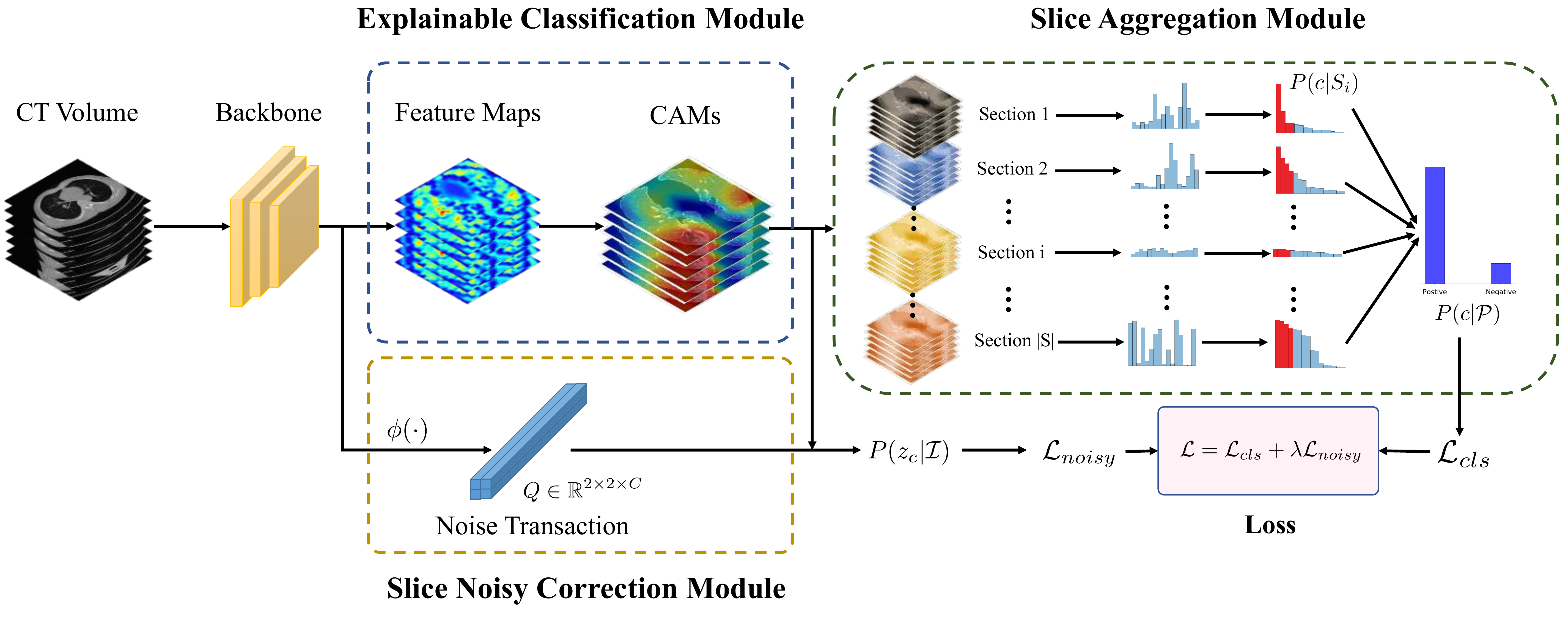}
\caption{The architecture of our proposed CIFD-Net. It is of note that $P(c \given S_i)$ denotes the probability of the Section $S_i$, and $P(c \given \mathcal{P})$ represents the probability of the patient who is tested COVID-19 positive or not. $Q\in \mathbb{R}^{2\times 2 \times C}$ indicates the noise transaction from the probability of the true label $P(y_c \given \mathcal{I})$ to the probability of the noise label $P(z_c \given \mathcal{I})$. In addition, $\phi(\cdot)$ is a feature embedding function. \revised{In addition, ResNet-50 \cite{he2016resnet} is adopted for backbone network.}}
\label{fig:network}
\end{figure*}

To cope with the issues above, in this work, we trained our model on both patient-level and image-level with multiple domain information. In particular, we consider the sequential information within the CT volume when predicting a patient is tested COVID-19 positive or not. To preserve the sequential information, we divide a lung CT volume into individual sections from the upper lobe all the way to the inferior lobe. As illustrated in Figure \ref{fig:network}, our method aggregates these sections as the representation of a patient. When aggregating these sections, we utilize the multiple instance learning method with the $k$-max selection strategy for images in each section. With the help of the $k$-max selection, our model can filter out the uncertain and noisy images that can be beneficial to make an accurate prediction. Moreover, multiple instance learning method is incorporated that can enforce our model to mine confident candidates for training and testing \cite{li2016weakly} while ignoring modeling the joint distribution of sections from the patient rather than a single image, which is rewarding for unseen center prediction.

In summary, our contributions are mainly three-fold:
\begin{itemize}
    \item We proposed a weakly supervised learning based multi-domain information fusion framework for automated COVID-19 diagnosis from multi-center and multi-scanner CT scans that only requires patient-level annotations for training.
    \item We propose a novel noisy label correction technique that propagates the patient-level predictions to individual slices and identifies the COVID-19 infected slices accurately.
    \item We develop a slice aggregation module to alleviate the data distribution shift problem,  which is essential for the deployment of the developed model in the real-world scenario. 
\end{itemize}

By validation on the China Consortium of Chest CT Image Investigation (CC-CCII) \cite{zhang2020ccii} benchmark dataset, our proposed coronavirus information fusion and diagnosis network achieves superior performance compared to state-of-the-art models on both patient-level and image-level.

\section{Related Work}

Before the COVID-19 pandemic, a huge amount of deep learning based methods has been proposed for lung cancer CT image analysis. In this research area, there have been great achievements, culminating in the development of many end-to-end pipelines for lung cancer diagnosis, classification, treatment planning, and prognostic evaluation \cite{ardila2019end,lakshmanaprabu2019optimal,hua2015computer,jiang2017automatic,hosny2018deep,setio2017validation,nishio2018computer}. In the treatment of interstitial lung disease (ILD), deep learning approaches have also been developed \cite{walsh2018deep,anthimopoulos2016lung,pang2019automatic,park2019lung}. In CT scans for COVID-19 patients, image characteristics, e.g., ground glass opacity and/or consolidation, are akin to those observed from lung cancer and ILD patient CT scans. Therefore, in the design of COVID-19 detection algorithms using CT images, insights from research on both lung cancer and ILD are significant and there is a clear translatability to the COVID-19 studies.


\paragraph{CNNs for Visual Recognition} 
Convolutional Neural Network (CNN) has been widely used in the medical diagnosis system \cite{ye2019dual, pham2018deep, wang2020covid}. Recently, plenty of COVID-19 recognition algorithms have been proposed, which have adopted artificial intelligence algorithms especially using the CNN. A comprehensive review of artificial intelligence assisted COVID-19 detection and diagnosis can be found elsewhere \cite{wynants_systematic_2020,shi_review_2020, yeCMBS2021, DBLP:journals/inffus/YangYX22, ma2021can}, and here we only provided a summary for the most relevant studies. 

Jin et al. \cite{jin2020ai} developed a combined segmentation-classification model for COVID-19 diagnosis. A few pre-trained models were tested, e.g., fully convolutional network (FCN-8s), U-Net, V-Net, and 3D U-Net++, as well as classification models like dual path network (DPN-92), Inception-v3, residual network (ResNet-50), and attention ResNet-50, from which the 3D U-Net++ and ResNet-50 combination achieved the best performance. However, it was unclear which layers were pre-trained and re-trained, the reproducibility of this study is uncertain. Wang and Wong \cite{wang2020covid} proposed COVID-Net, which stacked multiple convolutional blocks with dilated convolution to recognize chest X-ray images. Li et al. \cite{li2020artificial} explored the patient label and used max-pooling strategy over features extracted by the CNN from a set of slices to make the prediction. In addition, Ouyang et al. \cite{ouyang2020dual} deployed a 3D CNN and used the residual learning mechanism to build the network, which incorporated the depth information of the CT volumes. Shan et al. \cite{shan2020lung} proposed a human-in-the-loop strategy for infection region quantification, in which a modified V-Net was developed incorporating bottleneck building blocks to reduce training costs. The human-in-the-loop training procedure output a segmentation for subsequent manual corrections performed by radiologists, and then these corrected data were input to re-train the network iteratively. More recently, Hu et al. \cite{hu2020weakly} proposed a weakly supervised multi-scale learning framework for COVID-19 classification and lesions detection, which demonstrated promising results but its performance may be hindered by using the patient-level labels that contain noise labeling.

For automatic prognostication of COVID-19 patients, Huang et al. \cite{huang2020serial} developed a two-step segmentation model that extracted lung and lobes region followed by pneumonia segmentation. Both steps used separated U-Net and at least two follow-up scans for each patient were analyzed. The authors found significant differences in lung opacification percentage between the initial and the first follow-up scans, but not between the first and the second follow-up scans. Although the study findings are intriguing, there are critiques on lacking important information essential to the reproducibility \cite{takahashi2020regarding}.

Although the aforementioned studies and many others have shown promising results \cite{murphy2020covid,ozturk2020automated,oh2020deep,apostolopoulos2020covid,butt2020deep,wang2020deep,chen2020deep,song2020deep,zheng2020deep,mei2020artificial,driggs2021machine,roberts2021common}, two major issues can prevent the widespread deployment of these methods: (1) most previously proposed approaches relied on heavily annotated ground truth, e.g., for the infectious areas and slice-based labeling and (2) domain-shift failure for multi-center and multi-scanned datasets and therefore, poor reproducibility was always a concern.

\paragraph{Multiple Instance Learning}
The multiple instance learning (MIL) is a weakly supervised learning problem that has been attempted in several studies including weakly supervised object localization \cite{li2016weakly}, video anomaly detection \cite{sultani2018real}, weakly supervised image segmentation \cite{xu2014weakly} and others. In the MIL framework, a bag can be defined as a set of instances or image slices. Positive bags are assumed to contain at least one instance from a certain category and negative bags do not contain any instances from that category. It is intuitive to consider the classification of CT volumes that contain multiple CT slices as a MIL problem. A few methods have been proposed to solve the MIL problem. For example, Oquab et al. \cite{oquab2015object} trained a CNN using the max-pooling MIL strategy to classify the object. However, some of the MIL pooling strategies, such as max-pooling and mean-pooling, very often lead to insufficient and unstable training because of gradient vanishing. To fix this problem, Ilse et al. \cite{ilse2018attention} combined the gated attention mechanism with the MIL strategy to solve the medical image classification problem, but it could not predict the instance label accurately. \revised{Chen et al. \cite{chen2019synergistic} developed a stylized generative method to transfer the knowledge from MRI to CT within unsupervised manner. Xia et al. \cite{xia2020uncertainty} utilized uncertainties along different volume angles to measure the importance of predicted labels. Chen et al. \cite{chen2021adaptive} modelled intra-consistency between two domains to align the feature distributions. However, these methods requires to train the model using both source domain and target domain, which cannot handle the unseen domain scenarios.} Our method will provide solutions to these limitations.

\paragraph{Domain Adaptation} 
Domain adaptation refers to the techniques aimed at improving the performance of machine learning tasks, e.g., classification, detection, segmentation, when training the classifier on the data only from the source domain, but testing it using related samples from a shifted target domain. Some approaches also use domain adaptation to help learn the feature representation. Hoffman et al. \cite{hoffman2014lsda} proposed a method that learned the difference between classification and detection tasks, and transferred this knowledge from the classifier to detectors using weakly annotated data. In addition, MIL was incorporated for learning feature representation and classifier \cite{hoffman2015detector}. Besides, Mahmood et al. \cite{mahmood2018deep} utilized transformations such as hue, saturation, contrast, and brightness for RGB images to change the color and texture of the images in the source domain. Existing domain adaptation methods tend to use strongly annotated data in the source domain in order to improve the recognition performance, while our methods will focus on a weakly supervised manner. In other words, our method will require no instance-level annotation or the auxiliary strongly annotated data for recognition.


\section{Proposed Method}
In this section, we introduce the proposed coronavirus information fusion and diagnosis network (CIFD-Net) with their key modules including an explainable classification module (ECM), a slice aggregation module (SAM), and a slice noisy correction module (SNCM), respectively as illustrated in Figure \ref{fig:network}. 

The proposed ECM integrates the generation of class activation mapping into the forward propagation of the CIFD-Net, enabling CAMs generation during training and testing, which provides explainable results for the prediction of our model. 

Besides, instead of training on image-level (slice-wise) labels, which requires a significant amount of labor for manual labeling, we propose the SAM to train on patient-level labels. We model the joint probability of slices for each patient by which slices are divided into several consecutive sections with equal length. We then compute the probability of each section by adopting a $k$-max selection strategy, which can ignore some slice with large uncertainty, thus reduce the noise during modeling the joint probability at the patient level. With the help of modeling the joint probability, our model pays more attention to modeling the distribution of affected sections leading to better generalization on multiple domains.

Moreover, in order to improve the accuracy on the image-level, we further proposed the SNCM, which models the transaction between the true label and noisy label since the labels at the patient-level are considered to be noisy with respect to slice-wise labels.

\subsection{Problem Formulation}
The ultimate goal of our model is to diagnose whether a patient is tested positive or negative given a 3D volumetric CT lung scan. Let $\mathcal{P} = [\mathcal{I}_1, \mathcal{I}_2, \cdots, \mathcal{I}_n]$ denotes the lung CT volume for a patient with $n$ CT slices, where $\mathcal{I}_i$ is a 2D CT slice image. Let $Y \in \mathbb{R}^{\{0,1\}}$ denotes whether a patient is tested to be COVID-19 positive or not. $Y = 1$ when the patient gets COVID-19, while $Y=0$ stands for the patient is not COVID-19 infected. During the training stage, we only have patient-level labels, and the number of CT lung slices can vary significantly.

\begin{figure}
\centering
\includegraphics[width=1\linewidth]{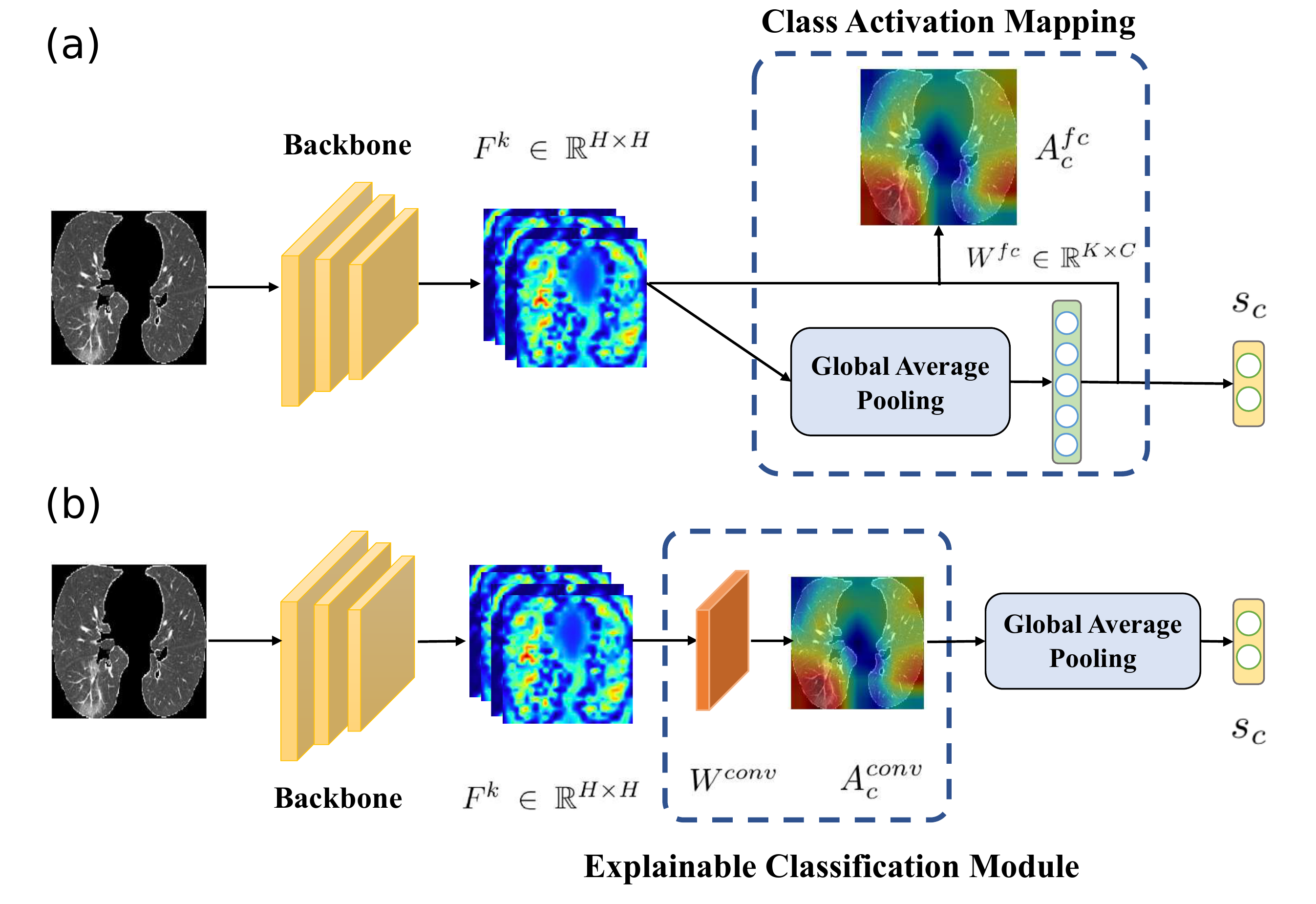}
\caption{(a) The workflow of the class activation mapping (CAM) scheme and (b) The proposed explainable classification module (ECM). It shows that our ECM can generate the CAM using only one forward pass, but the original method proposed by Zhou et al. \cite{zhou2016cam} needs a post-processing procedure to generate the CAM. \revised{$F^k$ is the $k$-th feature map from the backbone network. Besides, $W^{fc}$ and $W^{conv}$ are the weights for the fully connected layer and the convolutional layer. $A^{fc}_c$ and $A^{conv}_c$ are the class activation maps for class $c$. $s_c$ is the class score for class $c$.}}
\label{fig:cam}
\end{figure}

\subsection{Explainable Classification Module}

As the predicting process of CNN is in a black box. Several techniques \cite{zhou2016cam, selvaraju2017gradcam} have been proposed to shed light on how CNN makes the prediction and obtains the remarkable localization ability without any supervision of localization maps. As an explainable auxiliary diagnosis tool for radiologists, we employ the class activation mapping (CAM) \cite{zhou2016cam}, which can generate the localization maps for the prediction from the output of the backbone networks, e.g., ResNet \cite{he2016resnet}, VGG \cite{simonyan2014vgg}, GoogLeNet \cite{szegedy2015googlenet}, etc. However, the process of generating CAM is a two-step process, in which the backbone network is trained on the dataset and utilizes the weights of the final fully connected layer to compute the weighted sum of feature maps of the last convolutional layer. Suppose $F^k \in \mathbb{R}^{H \times W}$ is the $k$-th feature map with height $H$ and width $W$ from the last convolutional layer, and $W^{fc} \in \mathbb{R}^{K\times C}$ is the weight of the last fully connected layer, where $C$ is the number of classes and $K$ is the number of feature maps from the last convolutional layer. Therefore, the class score $s_c$ of the class $c$ can be calculated by

\begin{align}
    s_c = \sum_{k=1}^{K} W_{k,c}^{fc} \left( \frac{1}{H\times W} \sum_{i = 1}^{H} \sum_{j=1}^{W} F^k_{i,j} \right).
\label{eq:cam_origin}
\end{align}

Therefore, the localization map for the class $c$ proposed in \cite{zhou2016cam} is defined by

\begin{align}
    A^{fc}_c = \sum_{k=1}^K W_{k,c}^{fc} F^k,
\end{align}

\noindent and we can visualize the object localization maps via $A^{fc}_c$. 

Although CAM is a useful way to locate the region, it requires a post-processing procedure to generate. In our method, we plug the generation of CAM into the network with only one forward pass. 
\revised{Instead of directly applying global average pooling after the last convolutional layer, we replace the fully connected layer using a $1\times 1$ convolutional layer with the stride of 1 before the global average pooling operation.}
Suppose the weight of the convolutional layer is $W^{conv} \in \mathbb{R}^{K\times C}$ which is the same mathematical form as the weight of the fully connected layer, i.e., $W^{fc}$, we tweak the Eq.(\ref{eq:cam_origin}) as follows,

\begin{align}
    s_c = \frac{1}{H\times W} \sum_{i = 1}^{H} \sum_{j=1}^{W} \left( \sum_{k=1}^{K} W_{k,c}^{conv} F^k_{i,j} \right),
\label{eq:class_score}
\end{align}

\noindent which results in the same output with Eq.(\ref{eq:cam_origin}). Thus, the modified CAM for the class $c$ is computed as

\begin{align}
    A^{conv}_c = \sum_{k=1}^K W_{k,c}^{conv} F^k.
\end{align}

The modified activation mapping can accurately indicate the importance of the activation from CT images and locate the infected areas of the COVID-19 patients, providing the explainable and reliable results for prediction. The region with higher activation score indicates more importance the activation contributed to the prediction. The  modified activation mapping can also offer the auxiliary diagnostic information for radiologists. The differences between the original CAM and our ECM strategy are demonstrated in Figure \ref{fig:cam}.

\subsection{Slice Aggregation Module}

In some mild COVID-19 cases, there might be only part of the CT volume that has an infection, and very often the lesions are quite small. If we simply treat all slices as COVID-19 positive and train a classifier with the image-level label, it could lead to a noisy learning and poor results as the consequence. To overcome this problem, we propose the SAM and use the joint distribution to model the probability of patient is COVID-19 positive or negative. We assume that lesions are consecutive and only affect adjacent slices, consequently, we use a section based strategy to tackle the problem. The intuition of using the section based strategy is that it can be directly mapped to the problem of multiple instance learning (MIL) \cite{zhou2004mil}. In MIL, samples are divided into two bags classified as positive and negative bags. A positive bag contains at least one positive instance and a negative bag only has the negative instance. In the problem, only bag labels (patient annotations) are provided, and sections can be treated as instances in the corresponding bags. 

Given a patient $\mathcal{P} = [\mathcal{I}_1, \mathcal{I}_2, \cdots, \mathcal{I}_n]$ with $n$ CT slices, we divide these slices into disjoint sections, which can be considered as a set that contains an equal number of consecutive CT slices, i.e., $\mathcal{P} = \{S_i\}_{i=1}^{|S|}$, where $|S|$ is the amount of sections for patient $\mathcal{P}$ as defined as follows,

\begin{align}
    |S| = \max \left(1, \left\lfloor \frac{n}{l_s} \right\rfloor \right),
\end{align}

\noindent where $l_s$ is an empirically designed parameter named as section length. 

Then the probability of patient $\mathcal{P}$ belonging to the class $c$ can be represented as

\begin{align}
    P(c \given \mathcal{P}) 
    &= 1 - \prod_{i=1}^{|S|} \left(1-P(c \given S_i)\right),
\label{eq:patient}
\end{align}

\noindent where $P(c \given S_i)$ is the probability of the $i$-th section $S_i$ that belongs to the class $c$. Instead of taking the average of each probability of the slice in that section, we take the $k$-max probability for each class to compute the section probability. This is because some slices may contain few infection regions which can confound the prediction. To alleviate this problem, we adopt the $k$-max selection method which can be formulated as

\begin{align}
    P(c \given S_i) = \sigma\left( \frac{1}{k} \max_{\substack{s^{(j)} \in M}} \sum_{j=1}^k s^{(j)}_c \right), \nonumber \\ s.t. \quad M \subset S_i, |M| = k.
\label{eq:section}
\end{align}

\noindent where $s^{(j)}_c$ is the top $j$-th class score of the slice in the $i$-th section for the class $c$, and $\sigma (x) = 1 / (1 + e^{-x})$ is the sigmoid function. Then we use the patient-level annotations $\textbf{y}$ as the ground-truth during the training. The classification loss can be formulated as

\begin{align}
    \mathcal{L}_{cls} = \sum_{c = 0}^1 \left[y_c\log P(c \given \mathcal{P}) + (1-y_c)\log (1-P(c \given \mathcal{P})) \right].
\label{eq:loss_cls}
\end{align}

\begin{table}
\caption{The number of CT samples used for training for each class collected by four different hospitals A, B, C, and D. Besides, details of the CC-CCII dataset are also listed, which was used in the independent testing stage. \revised{The ratio of positive and negative samples in training set is approximately 1:1, and 2:1 in test dataset.}}
\begin{center}
\resizebox{0.968\linewidth}{!}{%
\begin{tabular}{c|ccc|ccc|c}
\hline
\multirow{2}{*}{\textbf{Dataset}} & \multicolumn{3}{c|}{\textbf{Number of Patients}}                  & \multicolumn{3}{c|}{\textbf{Number of CT Images}}                     & \multirow{2}{*}{\textbf{Subset}} \\ \cline{2-7}
                                   & \textbf{Total} & \textbf{Positive} & \textbf{Negative} & \textbf{Total} & \textbf{Positive} & \textbf{Negative} &                                  \\ \hline
Hospital A                         & 424           & 0                 & 424              & 24,670         & 0                 & 24,670            & Train                            \\
Hospital B                         & 58             & 58                & 0                 & 5,512           & 5,512              & 0                 & Train                            \\
Hospital C                         & 17           & 17              & 0                 & 2,611         & 2,611            & 0                 & Train                            \\
Hospital D                         & 305           & 305              & 0                 & 12,374         & 12,374            & 0                 & Train                            \\ \hline
CC-CCII \cite{zhang2020ccii}                            & 2,034           & 1,320              & 714               & 130,511         & 84,629            & 45,882             & Test                             \\ \hline
\end{tabular}
}
\end{center}
\label{tb:dataset}
\end{table}

\subsection{Slice Noisy Correction Module}
To further alleviate the negative impact of the image-level noises, we propose the SNCM, which is loosely inspired by \cite{bekker2016training}, to model the hidden distribution $P(z_c = i \given y_c = j, \mathcal{I})$ between the noisy label and the true label. Let $P(y_c \given \mathcal{I})$ denotes the true posterior distribution, given an image $\mathcal{I}$. The distribution of noisy label, $P(z_c \given \mathcal{I})$, can be modeled as

\begin{align}
    P(z_c = i \given \mathcal{I}) = \sum_j P(z_c = i \given y_c = j, \mathcal{I}) P(y_c = j \given \mathcal{I}).
\label{eq:noise_prob_simple}
\end{align}

We estimate the noise transaction $Q^c_{ij} = P(z_c = i \given y_c = j, \mathcal{I})$ for the class $c$ as follows

\begin{align}
    Q^c_{ij} = P(z_c = i \given y_c = j, \mathcal{I}) = \frac{\exp(w^c_{ij} \phi(\mathcal{I}) + b^c_{ij})}{\sum_i\exp(w^c_{ij} \phi(\mathcal{I}) + b^c_{ij})},
\label{eq:noise_trans}
\end{align}

\noindent where $i, j \in \{0,1\}$; $\phi(\cdot)$ is a nonlinear mapping function; $w^c_{ij}$ and $b^c_{ij}$ are trainable parameters for the class $c$ between the status $i$ and $j$. Transaction score $T^c_{ij} = w^c_{ij} \phi(\mathcal{I}) + b^c_{ij}$ can be regarded as the score of the transaction from the true label $i$ to the noisy label $j$ with respect to the class $c$. As a result, the estimated probability of noisy label for the class $c$ is represented as

\begin{align}
    P(z_c = i \given \mathcal{I}) = \sum_j Q^c_{ij} P(y_c = j \given \mathcal{I}).
\label{eq:noise_prob}
\end{align}

Finally, with the help of the estimated noisy probability, for the patient $\mathcal{P}$, the noisy classification loss is computed by

\begin{align}
    \mathcal{L}_{noisy} = \frac{1}{N} \sum_{i=1}^N\sum_{c = 0}^1 [y_c\log P(z_c = 1 \given \mathcal{I}) \nonumber \\+ (1-y_c)\log P(z_c = 0 \given \mathcal{I})].
\label{eq:loss_noisy}
\end{align}

By combining Eq. (\ref{eq:loss_cls}) and Eq. (\ref{eq:loss_noisy}), we can obtain the total loss function that we need to optimize for our model that is calculated as follows,

\begin{align}
    \mathcal{L} = \mathcal{L}_{cls} + \lambda \mathcal{L}_{noisy},
\label{eq:loss}
\end{align}

\noindent where $\lambda$ is a hyper-parameter to balance the loss terms. 

During the model training, the above loss functions are optimized iteratively. By incorporating the SAM, we can build a unified end-to-end deep neural network architecture for the COVID-19 diagnosis. The whole training procedure is summarized in Algorithm \ref{algo:train}.






\begin{algorithm}[!t]
\begin{algorithmic}[1]
\caption{Training procedure of CIFD-Net}
\label{algo:train}

\REQUIRE ~~\\
Set of $M$ CT volumes $\mathcal{P} = \{\mathcal{P}_1, \mathcal{P}_2, \cdots, \mathcal{P}_M\}$. \\
Learning rate: $\eta$.
\ENSURE ~~\\
Learned model parameters of CIFD-Net: $\Theta$.\\

\STATE Initialize all parameters denoted $\Theta$ using Xavier.
\REPEAT
	\FOR{$m = 1$ \TO $M$}
		\STATE Use the backbone network to compute slice features for CT volume $\mathcal{P}_m$ with its slices $\{\mathcal{I}_i\}_{i=1}^n$.
		\STATE Calculate class scores $s_c \in \mathbb{R}$ using Eq.(\ref{eq:class_score}).
		\STATE Compute the probability of patient $P(c \given \mathcal{P}_m)$ via SAM in Eq.(\ref{eq:patient}) and Eq.(\ref{eq:section}), and obtain classification loss $\mathcal{L}_{cls}$ via Eq.(\ref{eq:loss_cls}).
		\STATE Formulate the noise transaction $Q_{ij}^c$ between the probability of noise label and that of the true label by Eq.(\ref{eq:noise_prob_simple}) and Eq.(\ref{eq:noise_trans}).
		\STATE Obtain the estimated probability of noisy label $P(z_c = i \given \mathcal{I})$ by Eq.(\ref{eq:noise_prob}), and calculate noisy classification loss $\mathcal{L}_{noisy}$ using Eq.(\ref{eq:loss_noisy}).
		\STATE Compute the final loss $\mathcal{L}$ with Eq.(\ref{eq:loss}).
		\STATE Update parameters: $\Theta \leftarrow \Theta - \eta \bigtriangledown_{\Theta} \mathcal{L}(\Theta)$.
	\ENDFOR
\UNTIL{convergence}

\RETURN $\Theta$.

\end{algorithmic}
\end{algorithm}

\section{Experiments and Discussions}
In this section, the effectiveness of our method is validated and the results are quantified. First, we provide some statistics of the datasets and describe the implementation details as well as the experimental settings, which are followed by the reported results, the ablation studies, and further discussions on the qualitative and quantitative results.

\subsection{Datasets} 
In order to verify the effectiveness of proposed model on the data from an independent hospital, we use data from several hospital, then test the model on an independent dataset. The datasets used in our study are summarized in Table \ref{tb:dataset}. We collect CT datasets from four different local hospitals and anonymize the data by removing all the patient identity information. In total, there are 804 CT scan volumes with 45,167 CT images, 380 of which are COVID-19 positive and the other 424 are negative cases. All COVID-19 positive cases are confirmed by the RT-PCR tests. We train on the cross-domain datasets collected from hospitals A, B, C, and D and test on an open public CC-CCII dataset \cite{zhang2020ccii}. CC-CCII dataset consists of 2,034 3D CT volumes with 130,511 CT images, which have been acquired by the CT scanner from a different manufacturer representing another image domain.

\subsection{Data Standardization, Pre-Processing}
Following the protocol used in \cite{zhang2020ccii}, we first normalized images with $z-score$ normalization, then we used the U-Net segmentation network \cite{ronneberger2015unet} to segment the CT images. After that, we randomly cropped a rectangular region whose aspect ratio is randomly sample in $[3/4, 4/3]$ and area randomly sampled in $[90\%, 100\%]$, then resized the region into $224 \times 224$ shape. Meanwhile, we randomly flipped the input volumes horizontally with 0.5 probability. The input data would be a set of CT volumes which are composed by consecutive CT slice images.

\subsection{Implementation Details}
We use ResNet-50 \cite{he2016resnet} as the backbone network pre-trained on ImageNet \cite{deng2009imagenet}. 
\revised{For data augmentation, we use random horizontal flipping for the input CT volume in the spatial dimension. Each image in a CT volume is randomly horizontal flipped with a probability of 0.5. Then, we resize them into the size of 224 $\times$ 224. In addition, brightness and contrast are randomly changed within the range [0.9, 1.1].}
The dropout rate is set to 0.7, $\lambda$ is set to $0.0001$, and the $L_2$ weight decay coefficient is set to $10^{-5}$.  During the training and testing stage,  we set $l_s = 16$ and $k = 8$ to compute the patient probability. We train our model using the Adam optimizer \cite{kingma2014adam} with the initial learning rate $\eta=1\times 10^{-3}$, and training is terminated after 4,000 iterations with a batch size 10. All experiments have been conducted on a workstation with 4 NVIDIA Tesla V100 GPUs using PyTorch.

\subsection{Quantitative Results}


\begin{table*}
\caption{Comparison results of our CIFD-Net method vs. state-of-the-art architectures on the CC-CCII dataset. \revised{* indicates the p-value $< 0.05$, and ** represents the p-value $< 0.01$.}}
\begin{center}
\resizebox{1.0\linewidth}{!}{%
\begin{tabular}{c|c|c|c|c|c|c|c}
\hline
\textbf{Annotation}            & \textbf{Method}                      & \textbf{Patient Acc. (\%)} & \textbf{Precision (\%)} & \textbf{Sensitivity (\%)} & \textbf{Specificity (\%)} & \revised{\textbf{$F_1$-score (\%)}} & \textbf{AUC (\%)} \\ \hline
\multirow{5}{*}{Patient-level} & ResNet-50 \cite{he2016resnet}        & 53.70$_{\pm0.02}^{**}$                 & 61.42$_{\pm0.08}^{**}$                   & 77.13$_{\pm0.10}^{**}$                     & 10.37$_{\pm0.17}^{**}$             & 68.38$_{\pm0.05}^{**}$    & 46.30$_{\pm0.10}^{**}$   \\
                               & COVID-Net \cite{wang2020covid}       & 53.62$_{\pm0.03}^{**}$                      & 61.35$_{\pm0.01}^{**}$                   & 77.18$_{\pm0.05}^{**}$                     & 10.06$_{\pm0.25}^{*}$ & 68.36$_{\pm0.02}^{**}$            & 44.53$_{\pm0.18}^{*}$      \\
                               & COVNet \cite{li2020artificial}       & 67.64$_{\pm0.04}^{**}$                      & 76.03$_{\pm0.08}^{*}$                   & 73.17$_{\pm0.07}^{**}$                     & 57.34$_{\pm0.18}^{*}$  & 74.57$_{\pm0.04}^{**}$            & 66.13$_{\pm0.15}^{*}$     \\
                               & VB-Net \cite{ouyang2020dual}         & 76.75$_{\pm0.04}^{*}$                      & 85.25$_{\pm0.10}^{**}$                   & 77.61$_{\pm0.07}^{**}$                     & 75.22$_{\pm0.19}^{*}$ & 81.25$_{\pm0.05}^{**}$            & 89.48$_{\pm0.16}^{*}$      \\
                               & CIFD-Net (Ours)                      & \textbf{89.25$_{\pm0.02}^{**}$}             & \textbf{89.98$_{\pm0.13}^{*}$}          & \textbf{93.86$_{\pm0.06}^{**}$}            & \textbf{80.67$_{\pm0.13}^{*}$}   & \textbf{91.91$_{\pm0.07}^{**}$}  & \textbf{93.22$_{\pm0.06}^{**}$}      \\ \hline
\multirow{4}{*}{Image-level}   & ResNet-50 \cite{he2016resnet}        & 67.29$_{\pm0.04}^{**}$                      & 68.23$_{\pm0.06}^{**}$                   & 92.95$_{\pm0.05}^{**}$                     & 20.40$_{\pm0.16}^{**}$    &   78.71$_{\pm0.02}^{**}$   &   53.43$_{\pm0.11}^{**}$              \\
                               & COVID-Net \cite{wang2020covid}       & 64.83$_{\pm0.08}^{*}$                      & 66.28$_{\pm0.07}^{*}$                   & \textbf{93.18$_{\pm0.02}^{**}$}            & 12.48$_{\pm0.04}^{**}$ &   77.46$_{\pm0.03}^{**}$    &   51.47$_{\pm0.09}^{**}$            \\
                               & COVNet \cite{li2020artificial}       & 70.79$_{\pm0.03}^{**}$                      & 83.09$_{\pm0.07}^{*}$                   & 68.95$_{\pm0.11}^{**}$                     & 74.10$_{\pm0.08}^{**}$ &   75.37$_{\pm0.05}^{**}$     &   73.08$_{\pm0.07}^{**}$            \\
                               & CIFD-Net (Ours)                      & \textbf{84.83$_{\pm0.02}^{**}$}            & \textbf{91.19$_{\pm0.03}^{**}$}          & 84.74$_{\pm0.07}^{**}$                     & \textbf{84.99$_{\pm0.11}^{**}$}  &  \textbf{87.86$_{\pm0.04}^{**}$} &  \textbf{89.63$_{\pm0.08}^{**}$}      \\ \hline
\end{tabular}
}
\end{center}
\label{tb:sota}
\end{table*}

\begin{figure}
\centering
\subfigure[Patient-level Annotation]{
\includegraphics[width=0.9\linewidth]{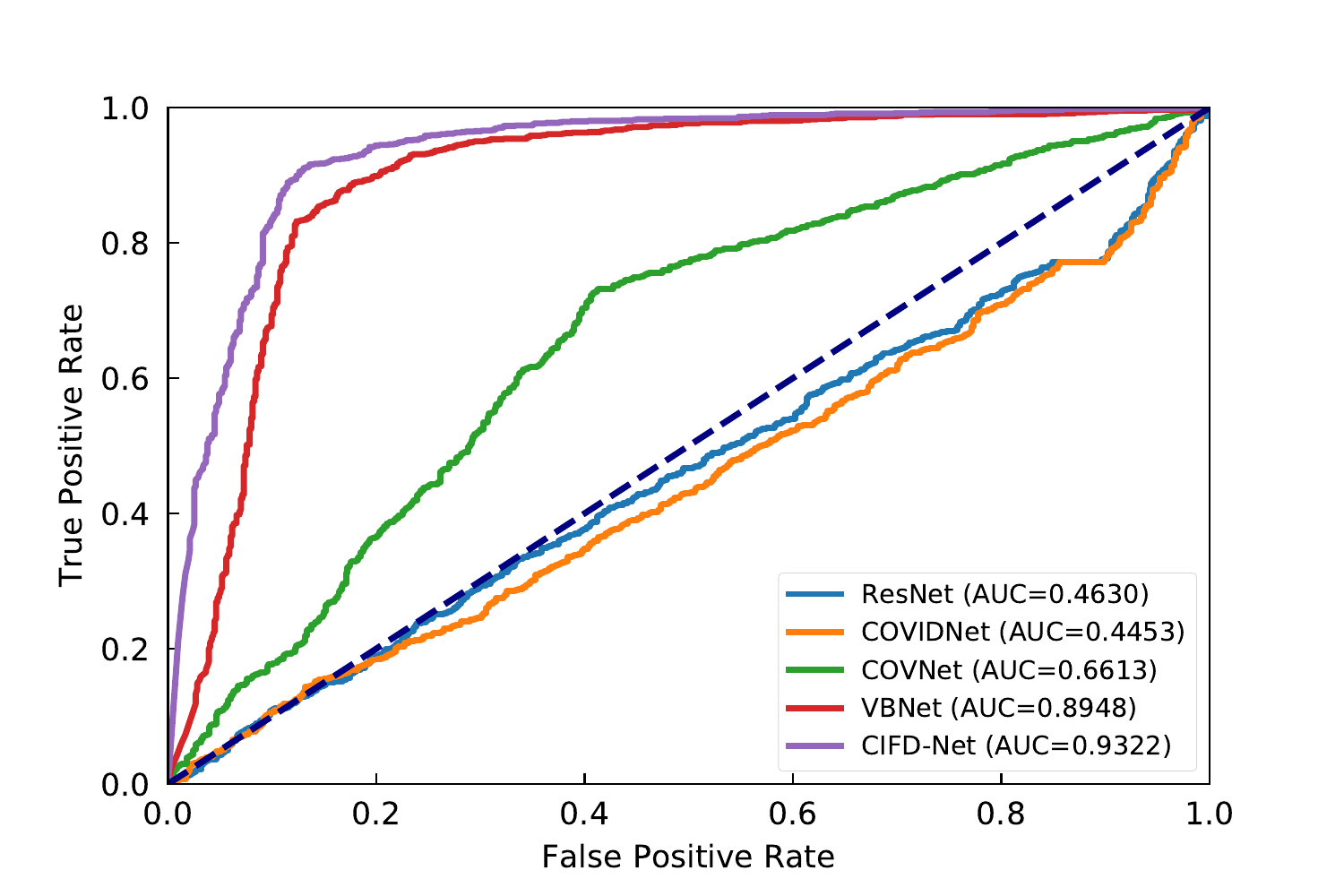}
}
\subfigure[Image-level Annotation]{
\includegraphics[width=0.9\linewidth]{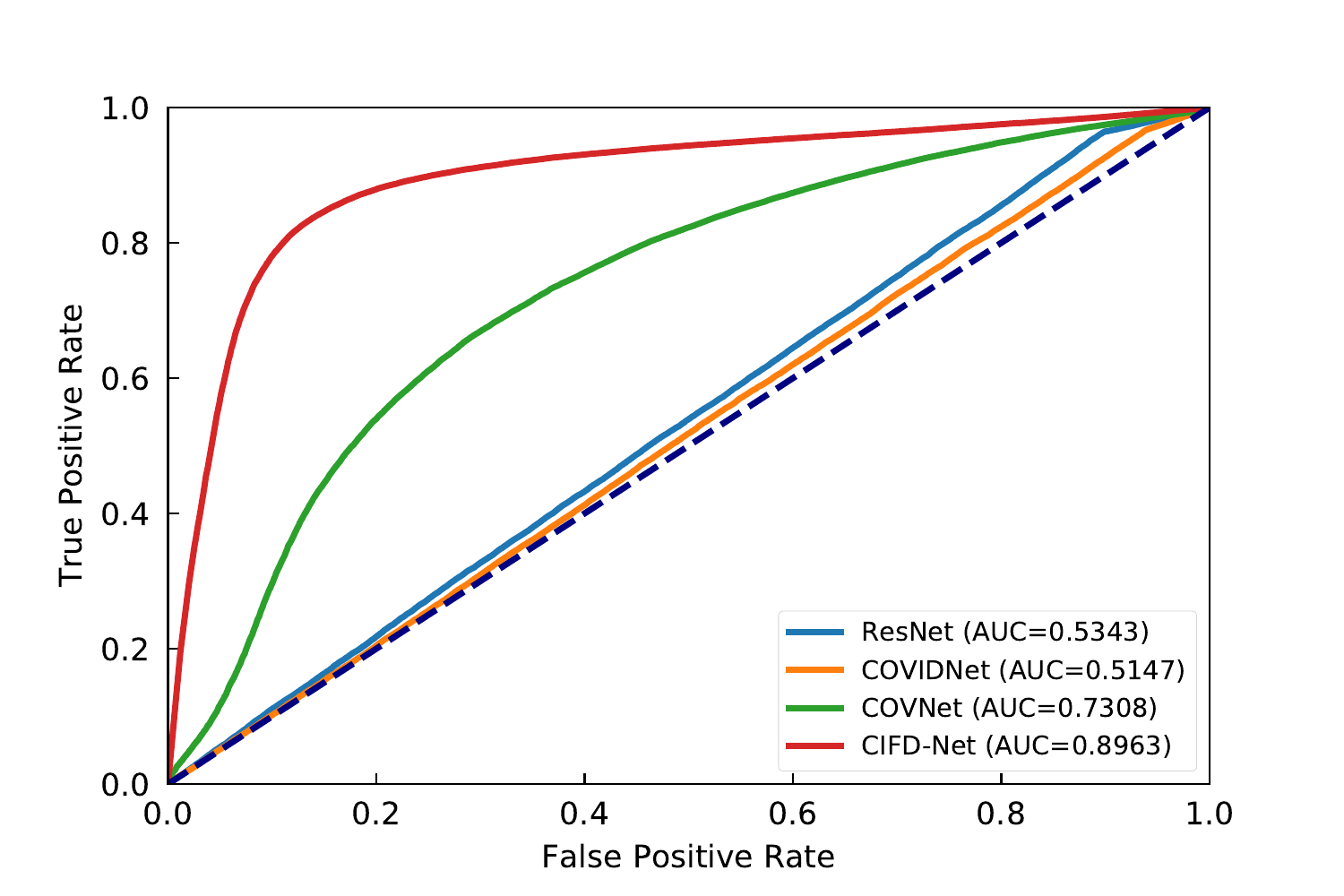}
}
\caption{Receiver Operating Characteristic (ROC) curves and area under ROC curves (AUC) of different models trained using patient-level annotation (a) and image-level annotation (b) on the CC-CCII dataset.}
\label{fig:roc}
\end{figure}

We reproduce and compare with four state-of-the-art methods \cite{he2016resnet, wang2020covid, li2020artificial, ouyang2020dual} on the COVID-19 CT classification. The results are shown in Table \ref{tb:sota}. For image-level supervision, COVID-Net \cite{wang2020covid} and ResNet-50 \cite{he2016resnet} employ the patient-level annotations as image annotations. Different to the methods proposed by Wang and Wong \cite{wang2020covid} and He et al.  and \cite{he2016resnet}, VBNet \cite{ouyang2020dual} adopts a 3D residual convolutional neural network (3D-ResNet) to train on CT volumes with patient labels. Moreover, COVNet \cite{li2020artificial} also trains on the patient-level label that they feed a patient-specific set of CT images into a 2D ResNet and simply aggregate the image-level feature descriptors with a max-pooling operator.~\\

From Table \ref{tb:sota}, several interesting observations can be summarized as follows.
\begin{itemize}
    \item The CIFD-Net outperforms most of the competing models by a large margin on the independent testing dataset, which can be attributed to the successful multi-domain shift problem proffered by our model. For the patient-level classification, our model is performed better than other compared methods by at least 12.5\% on accuracy. Moreover, our model yields also the best performance on the image-level classification, which outperforms COVNet \cite{li2020artificial} by 14.1\%. In addition, receiver operating characteristic (ROC) analysis and area under curves (AUC) results are obtained to quantify the classification performance. Our CIFD-Net achieves higher AUC value at both patient-level and image-level annotation compared to other state-of-the-art methods. Meanwhile, it is worth noticing that at the patient-level our method significantly outperforms other methods by at least 16.3\% with respect to the sensitivity, which is an important indication for diagnosing COVID-19 positive cases.
    \item Models trained on patient-level, such as \cite{li2020artificial}, \cite{ouyang2020dual} and ours, achieve significant performance improvement than those trained on the image-level, i.e., \cite{he2016resnet} and \cite{wang2020covid}, especially on the patient-level accuracy. This reflects that the image-level noise is non-trivial and can have a negative impact that these models can be overfitted because of the noise. Moreover, the models trained on the image-level may rely on learning the image textures \cite{geirhos2018imagenet}, which are highly discriminative between domains. As a consequence, the models are prone to be overfitted and biased toward different textures while predicting, which may explain why these methods, e.g., methods proposed in \cite{he2016resnet} and \cite{wang2020covid}, are poorly generalized to the unseen domains. 
    \item Although methods proposed by Li et al. and Ouyang et al. \cite{li2020artificial} and \cite{ouyang2020dual} also trained on the patient-level labels, our proposed CIFD-Net is superior to these methods, especially on the patient-level classification. The method proposed by Li et al. \cite{li2020artificial} performed the worst and this may because it has been trained on randomly selected CT images extracted from each 3D volume that may impede the encoding of lesions (often appearing adjacently between slices). In contrast, Ouyang et al. \cite{ouyang2020dual} preserved the sequential information among the CT slices because their method was trained on the whole CT volumes. In contrast, we take the full 3D volume into account and preserve the sequential information by dividing the volume into sections \cite{li2020artificial,ouyang2020dual}. Besides, VB-Net achieves better performance than COVNet because VB-Net is trained with stronger supervision that is additional to the image level classifier. It also employs an auxiliary pixel-wise classifier trained with pixel-level infection annotation (i.e., infection segmentation mask). In comparison, our proposed model achieves better overall classification performance than VB-Net with weak supervision only.
\end{itemize}

We carried out the ROC analysis and the AUC results were used to quantify the classification performances as shown in Figure \ref{fig:roc}. From Figure \ref{fig:roc} (a), we can observe that the models trained only on image-level annotations (i.e., ResNet-50 and COVID-Net) are not reliable since their AUCs are less than 50\%. In addition, we found that overall our CIFD-Net remains the best performed algorithm with an AUC of 93.22\%. It is of note that the overall results at the patient-level are higher than those at the image-level. This could be correlated with our findings in the classification that some CT slices with few lesion parts are hard to diagnose and classify.

\begin{table}
\caption{Accuracy (\%) of all the cases where each proposed component is applied.}
\begin{center}
\resizebox{1\linewidth}{!}{%
\begin{tabular}{c|ccc|c|c}
\hline
\textbf{Exp.} & \multicolumn{1}{c}{\textbf{ResNet-50}} & \textbf{$\bf{\mathcal{L}_{cls}}$} & \textbf{$\bf{\mathcal{L}_{noisy}}$} & \textbf{Patient Acc. (\%)} & \textbf{Image Acc. (\%)} \\ \hline
1            & $\surd$                                &                           &                               & 53.72                      & 67.31                    \\
2            & $\surd$                                & $\surd$                   &                               & 83.97                      & 78.60                    \\
3            & $\surd$                                &                           & $\surd$                       & 35.10                      & 35.16                    \\
4            & $\surd$                                & $\surd$                   & $\surd$                       & \textbf{89.23}             & \textbf{84.83}           \\ \hline
\end{tabular}
}
\end{center}
\label{tb:ablation_loss}
\end{table}

To examine the influence of different loss terms, we conduct ablation studies on the proposed model and the results are reported in Table \ref{tb:ablation_loss}. As seen in the table, the model with the SNCM slightly outperforms the model without the SNCM on the patient-level. However, the SNCM advances the prediction at the image-level with significant improvement by 6.2\% for the image accuracy. However, when only use the SNCM, the model would still be biased to predicting CT images tested negative because we only require our model to correct those CT images wrongly labeled as COVID-19 positive providing strong prior information to the training procedure.

\begin{figure}
\centering
\includegraphics[width=1\linewidth]{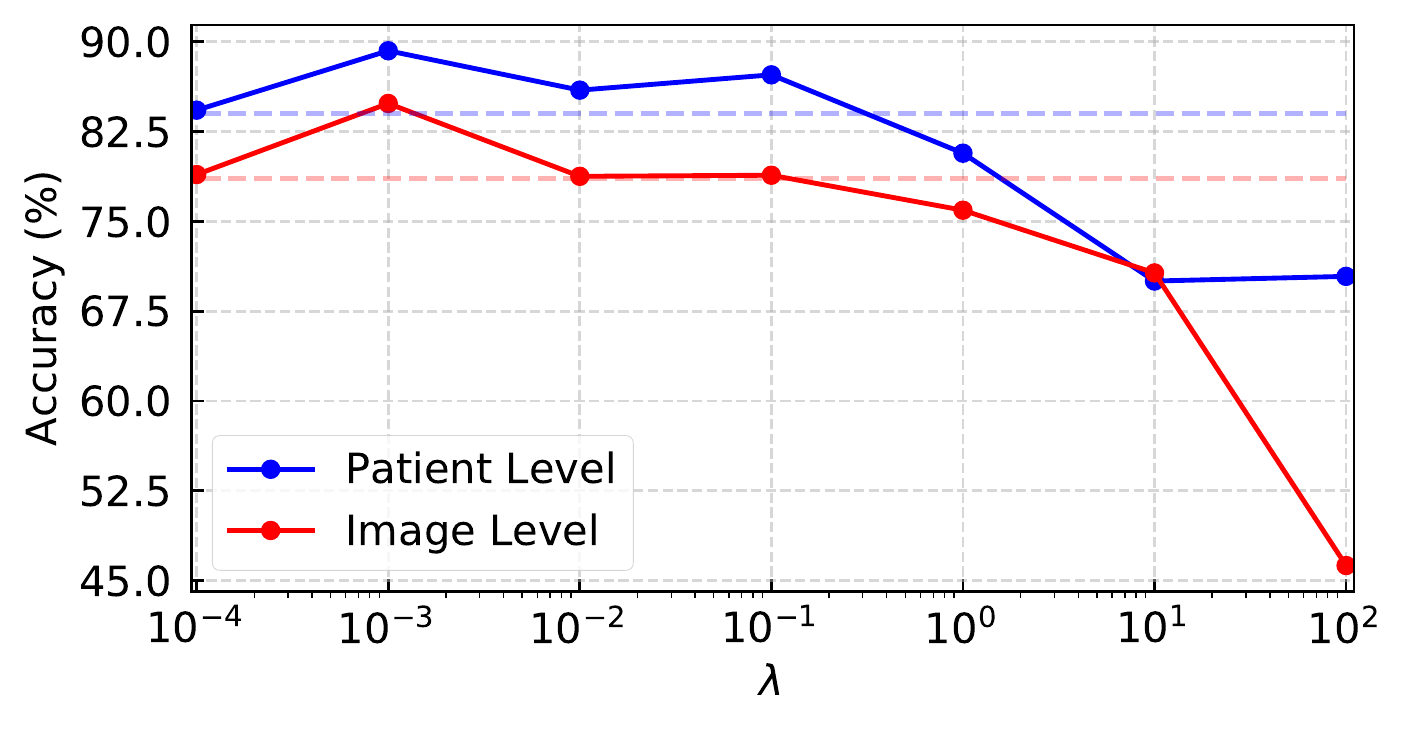}
\caption{Variations in classification results by changing the hyper-parameter $\lambda$. The light dash line represents the case when $\lambda = 0$. It shows that our model achieves the best performance with $\lambda = 1\times10^{-3}$.}
\label{fig:alpha}
\end{figure}

\begin{figure}
\centering
\includegraphics[width=1\linewidth]{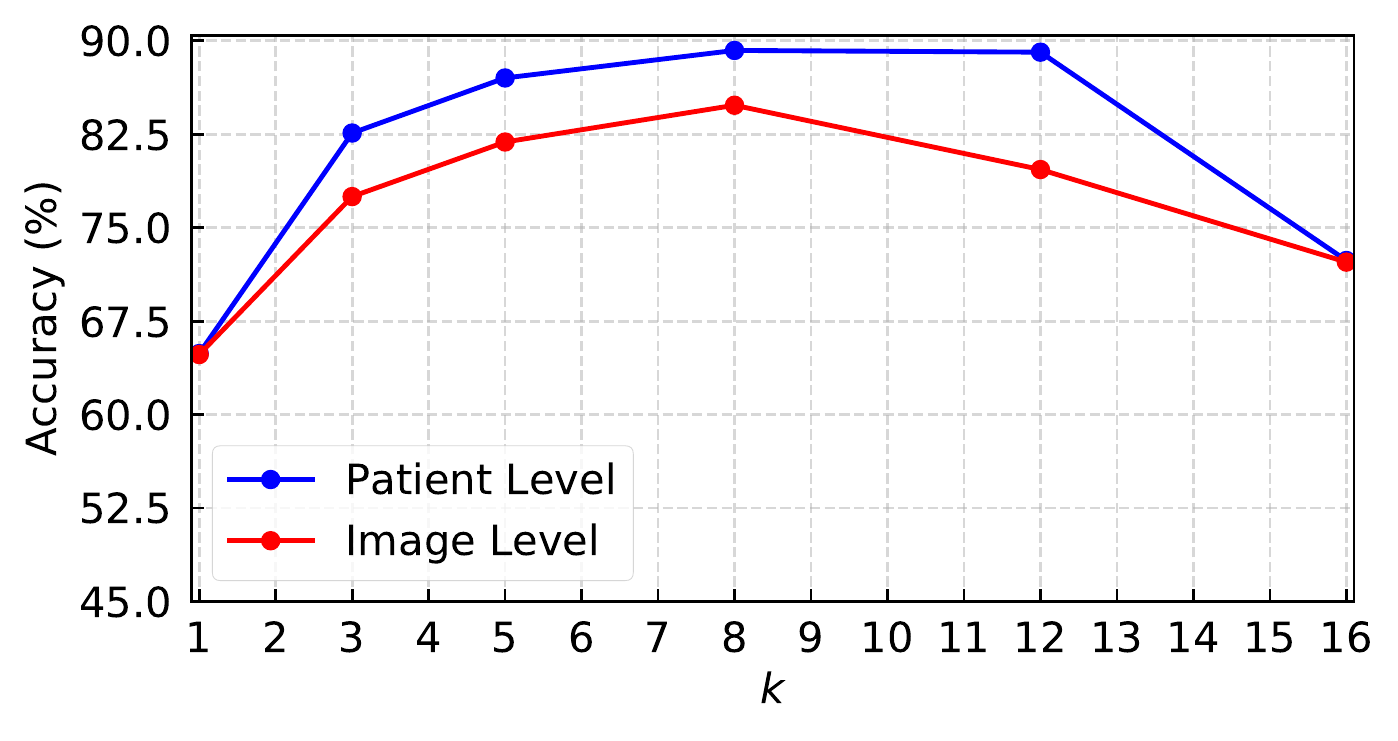}
\caption{Variations in classification results by changing the hyper-parameter $k$. Our model achieves the best performance with $k = 8$ with section size $l_s = 16$.}
\label{fig:topk}
\end{figure}


\begin{figure}[!ht]
\centering
\includegraphics[width=1\textwidth]{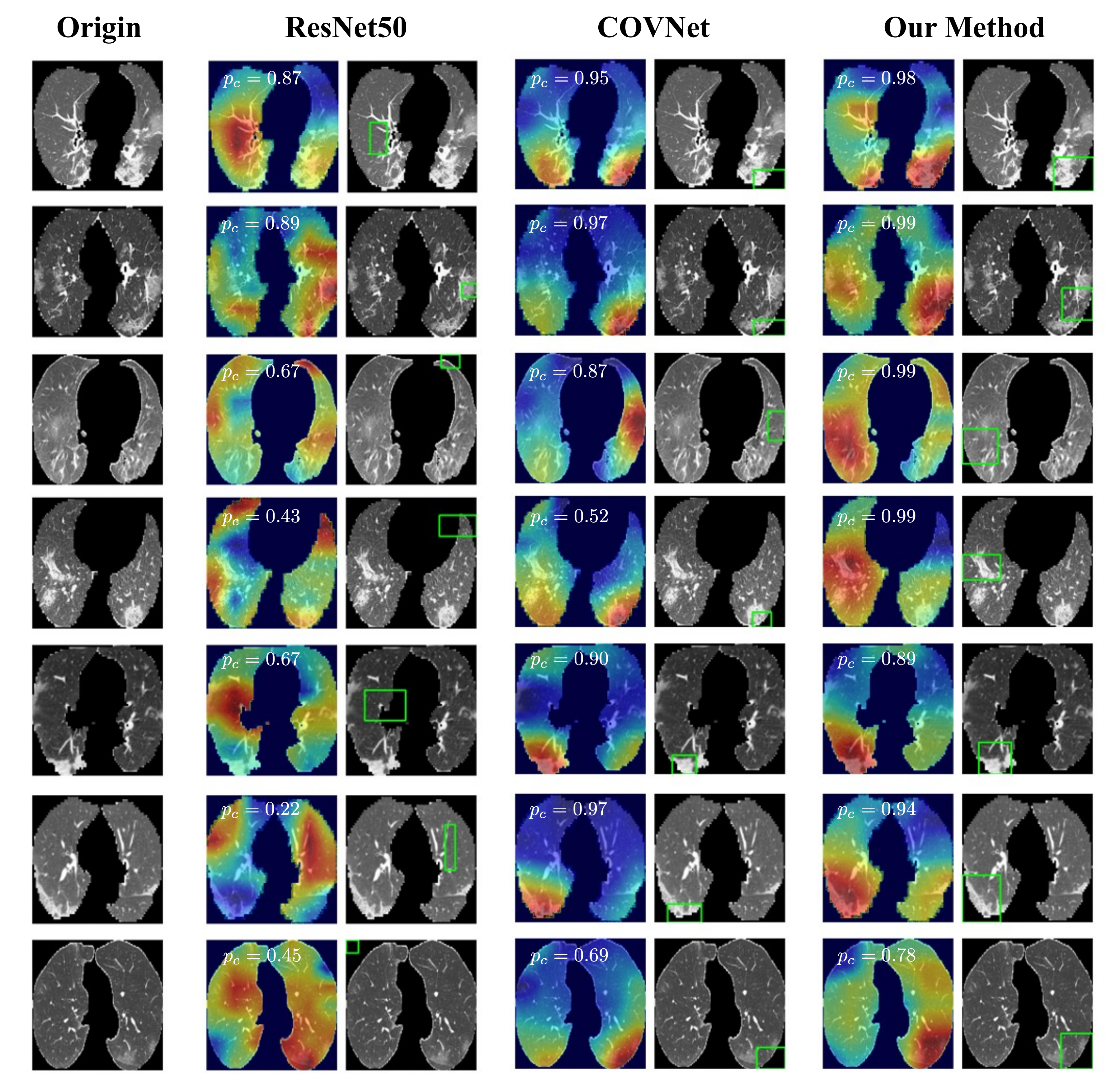}
\caption{Visualization of the CAMs and bounding boxes generated by different methods on the CC-CCII dataset. The region with a deeper red color indicates discriminative regions for the prediction by the model. \revised{$p_c$ is the probability for being predicted as COVID-19 positive.}}
\label{fig:cam_comparison}
\end{figure}


Furthermore, we have examined the sensitivity of the choice of the hyper-parameters $\lambda$ and $k$ for our model. Figure \ref{fig:alpha} shows the effect of the patient-level accuracy and the image-level accuracy while tuning the hyper-parameter $\lambda$. We can see that if $\lambda$ is too large, the model would be biased and the performance would drop significantly since it acts as the regularization terms in the model training. The best results are obtained when $\lambda = 1\times 10^{-3}$. In addition, for the selection of the hyper-parameter $k$, we can observe that when $k$ is too large or too small, the performance degrades dramatically. This is because that if $k$ is too large, the uncertainty of the section would increase and cause the noisy prediction. On the contrary, if $k$ is too small (e.g., $k = 1$), some important slice information would be neglected, which leads to inaccurate results.

\begin{figure}
\centering
\includegraphics[width=0.42\linewidth]{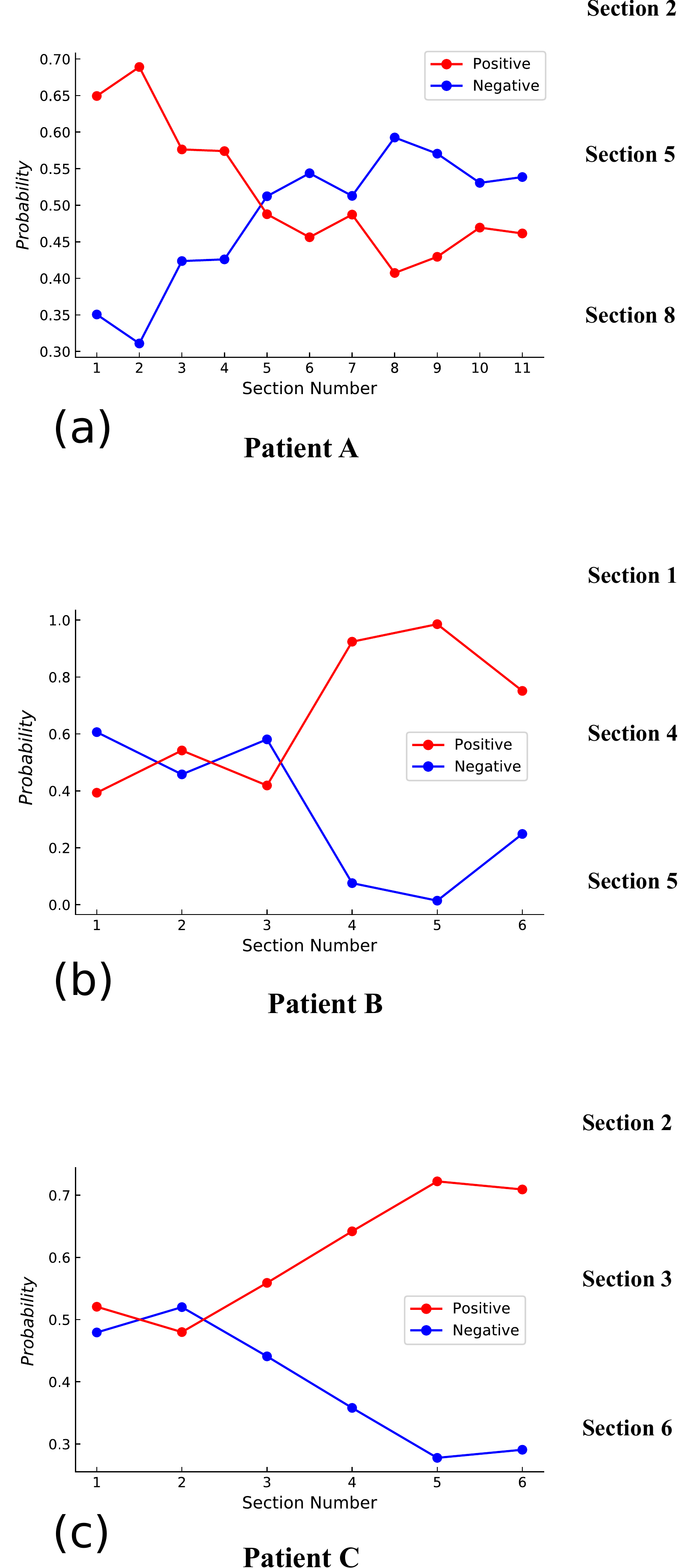}
\caption{Visualizations of infected/non-infected probabilities of each section for the patients. The x-axis of the plot is the section index of the patient. The right sub-figures of the probability plot are the picture sampled from the section listed above. (a) The first few sections are recognized as COVID-19 positive with high probability and when approaching the last few sections, no obvious lesions are found thus the positive probability drops drastically. (b) It shows that the probabilities of the first three sections are close to 0.5 indicating uncertainty for these sections. (c) For the last few sections, the lesions are gradually showing up in the left and right lower lobes together with increased infected probability.}
\label{fig:section}
\end{figure}

\subsection{Qualitative Results}
For qualitative studies, we use the trained models (e.g., ResNet-50, COVNet, and others) to visualize the CAMs and bounding boxes on the test set. Figure \ref{fig:cam_comparison} presents the visualization of CAMs using our ECM. We can clearly see that the model trained on the slice-level (ResNet-50) tend to discard the lesions and focus on non-infected regions, and this also explains why it makes inaccurate and unreliable diagnosis decision causing trouble for radiologist use. On the contrary, models trained on patient-level, COVNet for instance, are able to detect some of the lesions occasionally but mostly failed in estimating the extent of the lesions reliably. In contrast, our model is not only precise in terms of lesion localization but also precise in estimating the extent of the infectious areas.

Moreover, based on the results of the CAMs, we extracted the bounding boxes using each method. It can be found that our CIFD-Net is able to yield more accurate bounding boxes on the salient part of the CAMs (Figure \ref{fig:cam_comparison}) comparing to other methods, which indicates that our methods can be more applicable to perform auxiliary diagnosis. For instance, in diffusive cases (Figure \ref{fig:cam_comparison} rows 1 to 4), our CIFD-Net method has produced more accurate saliency maps compared to ResNet-50 and COVNet with less false positives and false negatives. Therefore, more precise localization (bounding boxes) have been generated. For the lesions distributed peripherally and subpleurally, both our CIFD-Net and COVNet approaches have performed better than the ResNet-50 (Figure \ref{fig:cam_comparison} rows 5 and 6). However, our CIFD-Net is more sensitive to the infectious regions that are not obvious in the images (Figure \ref{fig:cam_comparison} row 7).

In addition, we visualize the infection probability of lung sections for patients and sample the CT slices from corresponding sections. As illustrated in Figure \ref{fig:section}, the red curve depicts the infection probability varying along different lung sections, and the blue curve, on the opposite, depicts the non-infection probability for each section. Overall, it can be seen that the infected lung sections are distributed adjacently and the transition between the sections is smooth. Besides, we found our model is capable and robust of localizing where the infected lung sections are, regardless of the scale and the types of lesions. For example, for patient A section 2, despite there is a very small lesion (GGO) peripherally, our model is still quite sensitive and is able to identify the infected section. Our model reaches around a saddle point, i.e., 0.5, when there are no apparent lesions detected, for instance, section 1 for patient B and section 2 for patient C.

\subsection{Discussions}

Our proposed CIFD-Net sequentially aggregates image-level features within a CT volume to alleviate the multi-domain shift problems, which turns out to be very effective and we have demonstrated that our CIFD-Net can be better generalized to unseen data domain compared to other state-of-the-art works. This may be attributed to 1) the $k$-max selection strategy: when optimizing the joint probability, only top-$k$ probabilities within each section have been considered. Besides, those confounded images are not considered, which can result in a robust prediction; 2) our loss function is designed for modeling the joint probability of the patient instead of the individual image slice. Compared with the naive models, e.g., plain ResNet-50 trained on single image slice, our model is less likely to overfit on varied image styles and appearance, e.g., due to assorted textures and contrasts of the images, because our model takes into account the relationship between sections and the correlation between images in each section. 

In addition, we integrated a novel slice noise correction module, i.e., SNCM, in the proposed CIFD-Net, which adds additional regularization to the optimization. Besides, we argue that this not only contributes to boosting the classification performance on the image-level prediction but also leads to more precise localization of lesions. However, since we trained the CIFD-Net under the assumption that CT slices are consecutive and lung segments (sections) are ordered, it may be difficult to handle disordered CT slices by using the slice aggregation, i.e., SAM and as a consequence, it may result in less accurate classification.  

\section{Conclusion}

In this study, we have proposed a robust COVID-19 recognition model named CIFD-Net, which exploits the ECM to assist radiologists for auxiliary diagnosis. To handle the volume information, the model adopts the SAM to combine different sections for the sake of modeling the joint probability of the patient is COVID-19 positive or not. In addition, we extend our CIFD-Net incorporating the SNCM to predict a single CT slice without any image-level annotations. To investigate the prediction performance of the proposed model, we conducted comprehensive experiments on publicly available CT datasets. Experimental results have verified the superiority of our model, which can solve the multi-domain shift problem efficiently and effectively, compared to other state-of-the-art methods.

\section*{Acknowledgment}

This work was supported in part by the European Research Council Innovative Medicines Initiative on Development of Therapeutics and Diagnostics Combatting Coronavirus Infections Award ‘DRAGON: rapiD and secuRe AI imaging based diaGnosis, stratification, fOllow-up, and preparedness for coronavirus paNdemics’ [H2020-JTI-IMI2 101005122], the AI for Health Imaging Award ‘CHAIMELEON: Accelerating the Lab to Market Transition of AI Tools for Cancer Management’ [H2020-SC1-FA-DTS-2019-1 952172], the Hangzhou Economic and Technological Development Area Strategical Grant [Imperial Institute of Advanced Technology], the British Heart Foundation [TG/18/5/34111, PG/16/78/32402], the UK Research and Innovation Future Leaders Fellowship (MR/V023799/1), the SABER project supported by Boehringer Ingelheim Ltd, the Project of Shenzhen International Cooperation Foundation (GJHZ20180926165402083), and the Clinical Research Project of Shenzhen Health and Family Planning Commission (SZLY2018018).

\bibliography{Covid_bib}

\end{document}